\documentclass[11pt]{article}

\usepackage[iso-8859-7]{inputenc}
\usepackage{epsfig}
\usepackage{graphicx}
\usepackage{epstopdf}
\usepackage{amsfonts}
\usepackage{amssymb}
\usepackage{amsthm}
\usepackage{amsmath}
\usepackage{multirow}

\newcommand{\newc}{\newcommand}
\newc{\ra}{\rightarrow}
\newc{\lra}{\leftrightarrow}
\newc{\be}{\begin{equation}}
\newc{\ee}{\end{equation}}
\newc{\bs}{\begin{split}}
\newc{\es}{\end{split}}
\newc{\ba}{\begin{eqnarray}}
\newc{\ea}{\end{eqnarray}}
\newc{\ov}{\overline}
\newc{\pa}{\partial}
\newc{\D}{\Delta}

\newc{\nn}{\nonumber}
\begin{document}
	\begin{titlepage}
		
		\vspace*{0.7cm}

		\begin{center}
			{	\bf 	Diphoton Resonance in F-theory inspired Flipped $SO(10)$  }
			\\[12mm]
			George~K.~Leontaris$^{a}$
			\footnote{E-mail: \texttt{leonta@uoi.gr}} and
			Qaisar~Shafi$^{b}$
			\footnote{E-mail: \texttt{shafi@bartol.udel.edu}}
			\\[-2mm]
			
		\end{center}
		\vspace*{0.50cm}
		\centerline{$^{a}$ \it
			Physics Department, Theory Division, Ioannina University,}
		\centerline{\it
			GR-45110 Ioannina, Greece }
		\vspace*{0.2cm}
		\centerline{$^{b}$ \it
		Bartol Research Institute, Department of Physics and Astronomy, University of Delaware,}
		\centerline{\it
			DE 19716,  Newark, USA}
		\vspace*{1.20cm}
		
		\begin{abstract}
			\noindent
	
	Motivated by the diphoton excess at 750 GeV reported by the ATLAS and CMS experiments, we
	present an F-theory inspired flipped $SO(10)$ model embedded in ${\cal E}_6$.
	 The low energy spectrum includes the three MSSM chiral families, vectorlike color triplets, 
	 several pairs of charged $SU(2)_L$ singlet fields $(E^c, \bar E^c)$, as well as MSSM singlets, 
	 one or more of which could contribute to the diphoton resonance. A total decay  width in the 
	 multi-GeV range can arise from couplings involving the singlet and MSSM fields.

		\end{abstract}
		
	\end{titlepage}
	
	\thispagestyle{empty}
	\vfill
	\newpage


\section{Introduction }

Recently, the LHC experiments  ATLAS and CMS,  have reported  an excess of  diphoton 
events at an invariant mass  around 750 GeV  from  data at LHC run 2 with $pp$ collisions 
at $\sqrt s= 13$ TeV \cite{atlas13,CMS:2015dxe}.  The absence of data  on other channels such as $WW,
 ZZ$ and $Z\gamma $ indicate that the interpretation of these events cannot be accommodated within  
  the Standard Model (SM). It has been suggested that an interpretation requires 
 new physics  beyond the known SM context. In fact this effect hints to the existence of singlets
  and vector like  states  in the low energy  spectrum of the theory. 
  The observed resonance could be explained by  a SM scalar or pseudoscalar
  singlet state $X$~\cite{Franceschini:2015kwy,Ellis:2015oso,Mambrini:2015wyu,Pilaftsis:2015ycr,Dutta:2015wqh} 
   with mass $M_X\sim 750$ GeV~\footnote{  Although the present experimental data are not sufficient 
      to unambiguously confirm the existence of the resonance event, 
  the discovery of similar resonances at some higher energy scale accessible by the  LHC is plausible. 
        In the present model we can, in principle, adjust the free parameters  to accommodate such resonances at
      higher mass scales.}. This state could be generated by 
  the gluon-gluon fusion mechanism while   subsequently it decays to two photons. 
  Schematically, this is described as follows
  \be  
  gg \to X\to \gamma\gamma \label{gXgamma}
   \ee   
In a renormalisable theory, the production and decay in this process can be realised through loops
involving appropriate vector-like  states. 
Remarkably, a common phenomenon in string theory  model building is the occurrence of new singlet fields
 and vector-like exotic states in the massless spectrum of the effective low energy models which
 can mediate such processes~\cite{Heckman:2015kqk,Cvetic:2015vit,Anchordoqui:2015jxc,Ibanez:2015uok,Palti:2016kew,
	Karozas:2016hcp,Cvetic:2016omj,Dutta:2016jqn,Faraggi:2016xnm,Anastasopoulos:2016cmg,
	Lazarides:2016ofd,Ito:2016zkz,Li:2016xcj,Kats:2016kuz,Badziak:2016cfd,Aparicio:2016iwr,Hamada:2016vwk,Ge:2016xcq}~\footnote{There 
	is a vast list of recent publications  on possible interpretations of the $\gamma\gamma$ excess.
	 For a comprehensive 	list of models and interpretations see~\cite{Staub:2016dxq}. }.  
 
 F-theory models  in particular  offer a wide range of possibilities\cite{Beasley:2008kw,Donagi:2008ca}. Unlike other string 
 constructions, they admit exceptional gauge symmetries such as ${\cal E}_8$ and its subgroups, which incorporate 
 naturally the concept of gauge coupling unification. Besides, when most of the successful old GUT groups are
  realised in an F-theory background they naturally predict vector-like pairs of quarks and leptons in the light spectrum.
In fact, when the GUT symmetry is $SO(10)$ or higher the appearance of such states is
 unavoidable~\cite{Beasley:2008kw}.

Inspired by the above facts, in this note we construct a flipped $SO(10)$ model embedded in 
  an F-theory motivated ${\cal E}_6$ unified gauge group~\cite{Gursey:1975ki,Achiman:1978vg,Shafi:1978gg}.
    We show that this construction includes  singlets 
  as well as vector-like states which come with the quantum numbers of SM particles capable 
  of mediating  processes such as the diphoton production. In addition, we  find that other 
  vector-like states  with exotic quantum numbers emerge from the adjoint decomposition.

  The study of this alternative embedding is well motivated in F-theory constructions where
  	the GUT symmetry can be as large as $E_8$.  Indeed,  in the restricted case 
  	of  minimal   $SU(5)$,   there is a unique  assignment of the hypercharge generator in this group. 
  	However, there are many possibilities with a larger GUT symmetry  and includes additional $U(1)$ factors. In the case of $SO(10)\to SU(5)\times U(1)_{X}$,  with the standard hypercharge assignment 	the extra $ U(1)_{X}$ factor is treated as a spectator, but there is no compelling reason for this. 
  	Similarly, in the $E_6$ case, there are two additional $U(1)$ factors that could contribute to
  	the hypercharge.   Thus, different embeddings lead to distinct phenomenological predictions.	
  	In this work we wish to consider an alternative embedding of the hypecharge generator and try to assess 
  	the model in terms of its low energy predictions.

    In order to obtain chiral matter we will assume the existence of a suitable four-form flux. 
     Of course, the flux depends on the choice of the four complex dimensional Calabi-Yau (CY)  manifold
     and the geometric properties of the divisor supporting the specific singularity
    (${\cal E}_6$ in the present case). For our present purposes however, we will work 
    in the spectral cover approach where the properties of our local construction 
     can be adequately described in the infinitesimal vicinity of the GUT divisor, and 
    therefore,  we will rely on the assumption that such a manifold exists.

 The layout  of the present paper is as follows. In the next section we 
 present an $SO(10)$ flipped model embedded in the ${\cal E}_6$  gauge symmetry.
 We discuss the  basic properties  of its spectrum and the predicted exotics. 
 In section 3 we derive the superpotential of the effective model emerging 
 under the action of a $Z_2$ monodromy. Next, in section 4 we  focus  on the existence 
 of exotic vector-like  pairs and singlet field which are suitable to  
 contribute to the diphoton emission in  $pp$ collisions.  We present our conclusions
 in section 5.

\section{Effective  flipped models from ${\cal E}_6$}

In F-theory  the gauge symmetry of the effective theory is linked to the
geometric singularity of the compactification manifold. In the elliptic 
fibration these singularities are described by the sequence 
of the subgroups of the exceptional group ${\cal E}_8$.    
In the present F-theory construction we will analyse an 
$SO(10)\times U(1)$ gauge symmetry which admits a natural embedding in the 
exceptional group ${\cal E}_6$. 
Therefore, with respect to the ${\cal E}_8$ we have  the following breaking pattern:
\[{\cal E}_8\supset{\cal E}_6\times SU(3)_{\perp}\supset SO(10)\times U(1)_{X'}\times SU(3)_{\perp}\]
where, in accordance to the standard terminology,
the  $SU(3)_{\perp}$ factor is  considered as the group  `perpendicular' to ${\cal E}_6$ GUT
divisor. We will  assume a semilocal  approach where
 the   ${\cal E}_6$ representations transform non-trivially under  $SU(3)_{\perp}$ .  The matter content arises from the 
 decomposition of the ${\cal E}_8$  adjoint (${\cal E}_8\supset {\cal E}_6\times SU(3)_{\perp}$)
\[248\ra (78,1)+(1,8)+(27,3)+(\ov{27}, \overline{3})\]
In the spectral cover approach the ${\cal E}_6$ representations 
are distinguised by the  `weights' $t_{1,2,3}$ of  the $SU(3)_{\perp}$ Cartan subalgebra
subject to $t_1+t_2+t_3=0$, while  the $SU(3)_{\perp}$  adjoint `decomposes'
into singlets $1_{t_i-t_j}\equiv \theta_{ij}$. We introduce the notation
\[
(1,8)\ra \theta_{ij},\;\
(27,3)\ra 27_{t_i},\;
(\ov{27},\overline{3})\ra\ov{27}_{-t_i}
\]
while the ${\cal E}_6$  adjoint ${\bf 78}$ is an  $SU(3)_{\perp}$ singlet and therefore carries no $t_i$ index. 
Since we are interested in a flipped $SO(10)$ model,  in the subsequent analysis we  choose
 to accommodate the ordinary fermionic states and Higgs in the ${27}_{t_i}$. We further 
 assume that the symmetry breaks though a non-trivial  abelian flux which, at the same time,  
 determines the chirality of the  complete ${\cal E}_6$ representations $27_{t_{1,2,3}}$.

We start with the derivation of the flipped $SO(10)$ model~\footnote{For previous work on 
	flipped $SO(10)$ see for 	example \cite{Maekawa:2003wm} and references therein.} in an F-theory inspired context.
We will assume that the bulk gauge group is ${\cal E}_6$, which breaks to $SO(10)$  by 
turning on a $U(1)_{X'}$ gauge field configuration, where the particular $U(1)_{X'}$ is 
embedded in ${\cal E}_6$. 
 Under the decomposition ${\cal E}_6\supset SO(10)\times U(1)_{X'}$,  
the relevant ${\cal E}_6$ representations decompose as follows
\ba 
{\cal E}_6&\supset& SO(10)\times U(1)_{X'}\nn\\
78&\ra& {\bf 45}_0+{\bf 1}_0+{\bf 16}_{-3}+\overline{\bf 16}_3\label{78}\\
27&\ra&{\bf 16}_1+{\bf 10}_{-2}+{\bf 1}_4\label{27}\\
\ov{27}&\ra&\ov{\bf 16}_{-1}+\ov{\bf 10}_{2}+{\bf 1}_{-4}\label{27n}
\ea
In principle, there are  $SO(10)$  zero modes in the adjoint ${\bf 16}_{-3}$ and $\overline{{\bf 16}}_3$
as well as in ${\bf 27}$, which might accommodate chiral matter provided that $n_{16}-n_{\overline{16}}\ne 0$. 
In the next step, we break the $SO(10)$ symmetry down to $SU(5)\times U(1)_X$ by turning on
a flux along $U(1)_{X}$, so that at this stage the symmetry breaking chain is
\be 
{\cal E}_6 \supset SO(10)\times U(1)_{X'} \supset  [SU(5)\times U(1)_X]\times U(1)_{X'}\,.
\ee
If we  denote with  $X,X'$ the corresponding abelian charges,
for the flipped $SO(10)$ case we define the  following combination~\footnote{For 
the $U(1)_{X',X}$ charge assignment, we adopt the convention 	${\bf 16}_1\to 10_{(-1,1)}+\bar 5_{(3,1)}+1_{(-5,1)}$,
${\bf 10}_{-2}\to 5_{(2,-2)}+\bar 5_{(-2,-2)}$ and $1_{(0,4)}$.}
\be 
Z=-\frac 14 \left(X+5 X'\right)\label{Zdef}
\ee 
 Under the above symmetry breaking, the $SO(10)$ representations decompose to various
 $SU(5)$ multiplets. With respect to  ${\cal E}_6\to SU(5)\times U(1)_Z$, these have 
 the following  `charge' assignments:
\ba
27&\ra&
\{10_{\bf -1}+\overline{5}_{\bf -2}+1_{\bf 0}\}+\{5_{\bf 2}+\bar 5_{\bf 3}\}+{\bf 1}_{\bf -5}\label{27dec}\\
78&\to& \{24_{\bf 0}+10_{\bf -1}+\ov{10}_{\bf 1}+1_{\bf 0}\} \nn\\
&&+\{10_{\bf 4}+\bar 5_{\bf 3}+1_{\bf 5}\}\nn\\
&&+\{\ov{10}_{\bf -4}+ 5_{\bf -3}+1_{\bf -5}\}\nn\\
&&+{\bf 1}_{\bf 0}\,\cdot\label{78dec}
\ea
At the final stage, we break $ SU(5)\to SU(3)\times SU(2)\times U(1)_y$, where the hypercharge
is defined to be the linear combination of the three abelian factors $U(1)_{X'}, U(1)_X, U(1)_y$ given by:
\be 
 Y =-\frac 15 \left(Z+\frac y6\right) \label{Zydef}
\ee 

 We will require  that the SM fermions and Higgs  doublets reside on matter curves $\Sigma_{27}$ formed at the intersections
	of the GUT surface with other 7-branes. Employing  the above hypercharge definition, the embedding of the 
	SM states in the 27-representation of $E_6$ is as follows:
\ba
{\bf 27}&=&\left\{\begin{array}{lll}{\bf 16}_{1}
	\ra
{\small \left(\begin{array}{cc}
         Q&d^c\\\nu^c&
         \end{array}\right)+\left(\begin{array}{c}
         \bar D\\h_u
         \end{array}\right)}+\phi\\
         {\bf 10}_{-2}\ra {\small \left(\begin{array}{c}
         D\\h_d
         \end{array}\right)+\left(\begin{array}{c}
         u^c\\ \ell
         \end{array}\right)}\\
         {\bf 1}_4\ra e^c
          \end{array}\right.\,,
         \ea

	The symbol $\phi$ stands for a $SU(5)$ singlet field while for all other SM states,
	we use  the standard notation.
	As can be seen, compared to the standard  $SO(10)$ embedding, here we obtain a `flipped' picture of the $5$-plet and singlet
	representations, i.e., $\bar 5_h\leftrightarrow \bar 5_f$ and $1_{e^c}\leftrightarrow 1_{\phi}$. 
	More precisely, compared  to flipped $SU(5)$, this  $U(1)_Z$ definition flips 
	$\bar 5_{-2}$ with $\bar 5_{3}$ and $1_0$ with $1_{-5}$.
	The fermion component $\bar f=\bar 5_{-3}$ in this case is part
	of the ${\bf 10}$-plet ($\in SO(10)$), and the Higgs $\bar h=\bar 5_{-2}$ is part of ${\bf 16}$. Furthermore,
	the $SU(5)$ singlet $\phi$ is electrically neutral and the right-handed electron is found in the $SO(10)$
	singlet ${\bf 1}_{4}$.
	
	In addition to the SM fields residing in the $\Sigma_{27}$ matter curves,    there is also bulk matter emerging from the 
	docomposition of the ${\bf 78}$ representation. Namely: 
\ba
{\bf 78}&=&\left\{\begin{array}{lll}
{\bf 45}_{0}\ra
G_0+T_0+S_0+Q+\overline{Q}\\
    \qquad +\left\{(Q+D^c+N^c)+c.c.\right\}+\chi_0\\
                  {\bf 16}_{-3}\ra
                  {\small \left(\begin{array}{cc}
                           Q'&U^c\\ \bar E^c&
                           \end{array}\right)+\left(\begin{array}{c}
                           U^c\\L
                           \end{array}\right)}+\bar E^c\\
    {\small \ov{\bf 16}_{+3}\ra
      {\small \left(\begin{array}{cc}
       \bar Q'&\bar U^c\\E^c&
       \end{array}\right)+\left(\begin{array}{c}
        \bar U^c\\\bar L
        \end{array}\right)}}+ E^c\\
         {\bf 1}_0\ra \Phi
          \end{array}\right.\,.\label{78com}
         \ea
We have used the symbols $\chi_0, \Phi$ for the two neutral singlets, 
while for the remaining  content arising from the decomposition 
of ${ 24}\in SU(5)$, we have introduced  the notation
\ba 
G_0+T_0+S_0&=& (8,1)_0+(1,3)_0+(1,1)_0\\
Q+\bar Q&=& (3,2)_{\frac 16}+(\bar 3,2)_{-\frac 16}\\
Q'+\bar Q'&=& (3,2)_{-\frac 56}+(\bar 3,2)_{\frac 56}\,.
\ea 
In the above,  $Q$ has the standard quark doublet quantum numbers and $\bar Q$ is its complex conjugate,
while $Q'+\bar Q'$ have exotic charges.

In the standard (non-flipped) $SU(5)$ theory, the $Q'+\bar Q'$ exotics emerge from the decomposition of
the $24$-adjoint. Hence,  we observe that the flipped case interchanges $Q'+\bar Q'$ exotics 
in the adjoint of the standard $SU(5)$,
with the ordinary $Q+\bar Q$ within the ${\bf 16}_{-3}+\ov{\bf 16}_{3}$ of $SO(10)$.

 If some of  these bulk states remain in the light spectrum, they could contribute
 to new physics phenomena with possible signatures in future experiments. We will comment on these issues 
	in the next section.

\section{Superpotential with $Z_2$ monodromy}

Having determined the particle spectrum of the effective field theory model, we 
 proceed now to the superpotential. We find it convenient to perform the
 analysis using the spectral cover approach. Given that there are two non-trivial ${\cal E}_6$
representations available, namely ${\bf 27}$ and ${\bf 78}$, the only possible tree level terms
are ${\bf 27}^3$, ${\bf 78}^3$ and ${\bf 78}\cdot {\bf 78}\cdot {\bf 1}_x$ where ${\bf 1}_x$ is a 
singlet embedded in the ${\cal E}_8$ adjoint.  We have explained in the introductory section 
that in the context of the  $SU(3)$ spectral cover
the fundamental representation is characterised by the corresponding  weights $t_i$ 
and the Yukawa couplings should respect the requirement $t_1+t_2+t_3=0$.
 Furthermore,  as is well known, a monodromy action is required to ensure 
 a top Yukawa coupling at the tree-level. We choose   a $Z_2$ monodromy, which identifies the 
 two weights $t_1=t_2$. We accommodate the fermion families in $27_{t_{1}}$ 
 and the Higgs in $27_{t_3}$,  so that a diagonal Yukawa term $27_{t_1}27_{t_1}27_{t_3}$ is allowed. 
 After the implementation of the $Z_2$ monodromy,  the condition for the weights $t_i$  becomes $2t_1+t_3=0$. 
 It is also worth observing that the spectral cover symmetry reduces 
 essentially to a $U(1)_q$  symmetry in the effective field theory model where the $U(1)_q$ charges 
 of the two matter curves are $t_1$ and $t_3=-2t_1$. Therefore, the symmetry of the effective model  is in fact
 \[SO(10)\times U(1)_Z\times U(1)_q\subset  {\cal E}_6\times U(1)_q\,. \]
 
 To define the homological properties of the matter curves, we recall that in the case of $SU(3)_{\perp}$ 
 	the spectral cover  is described by a cubic polynomial whose roots  the $t_i$.
 For the  case of $Z_2$ monodromy we assume the factorisation~\footnote{Note that,  because
 	the sum of the roots is zero,  $b_1\equiv\sum _it_i=0$.}
 	\be 
 	 b_0s^3+b_2s+b_3 = (a_1+a_2s+a_3s^2) (a_4+a_5 s)\,,\label{sce}
 	\ee
 	where the $b_k$'s homologies  are $[b_k]=\eta-kc_1$ and $[s]=c_1$.
 	Here $\eta=6c_1-t$, where $c_1=c_1(S)$ is the first Chern class of the GUT ``surface'' $S$ and,
 	$c_1(N^{\perp})=-t$ that of  the normal bundle.
 	
	 The second degree polynomial of the right part of the above equation means that 
  two roots are not separable within the field of holomorphic functions, and as a result, a  $Z_2$ monodromy 
  identifies the two weights $t_{1}=t_2$ in accordance with our assumptions stated above. 
 Moreover, equation (\ref{sce}) implies the following relations $b_k=b_k(a_i)$ between  the 
 coefficients
  \be b_0=a_3a_5, \, b_1=a_2a_5+a_3a_4=0,\, b_2=a_1a_5+a_2a_4,\, b_3=a_1a_4\,,
  \label{bacoefs}
  \ee
   which can be used to determine the homologies of $a_i$'s.  
   Furhermore, the  equation  $b_3=a_1a_4=0$ of the  ${\bf 27}$, implies that the 
   two 	matter curves  $27_{t_1}$ and $27_{t_3}$  are associated with the defining equations $a_1=0$ and $a_4=0$ 
   respectively. 
   
   From~(\ref{bacoefs}), we infer that the homologies satisfy relations of the form $[b_k]=[a_l]+[a_{8-l-k}]$
so that   it can be readily found~\cite{Callaghan:2011jj,Callaghan:2013kaa} that the  
   $27_{t_1}$ and $27_{t_3}$ homologies are $\eta-2c_1-\chi$   and $\chi-c_1$ respectively where 
   $\chi$ is left unspecified. Then, assuming a $U(1)$ flux piercing these matter 
   curves,  the multiplicities of $27_{t_i}-\overline{27}_{t_i}$ are
   given by the restrictions   $n_1={\cal F}_{U(1)}\cdot (\eta-2c_1-\chi)$ and $n_3={\cal F}_{U(1)}\cdot (\chi-c_1)$
   (with ${\cal F}_{U(1)}$ denoting the abelian flux). From this, we deduce that the chiral states of 
   the model are given by
   \[  n_1+n_3 = {\cal F}_{U(1)}\cdot (\eta-3c_1)\equiv{\cal F}_{U(1)}\cdot (3c_1-t)\,,  \]
and therefore, the unknown homology $\chi$ does not play any r\^ole in the determination of the chiral spectrum.
Hence, to obtain three chiral families we impose $n_1+n_3= {\cal F}_{U(1)}\cdot (3c_1-t)=3$.
\begin{table}
	\begin{center}
		\begin{tabular}{cclll}
			Matter & Equation & Homology & $\# 27_{t_i}-\# \overline{27}_{-t_i}$ \\
			\hline
			$27_{t_{1}}$ &  $a_1$ & $\eta - 2 c_1 - {\chi}$  & $n_1={\cal F}_{U(1)}\cdot (\eta-2c_1-\chi)$ \\
			$27_{t_{3}}$ &  $a_4$ & $-c_1 + \chi$ & $n_3={\cal F}_{U(1)}\cdot (\chi-c_1)$ \\
		\end{tabular}
		\caption{${\cal E}_6$ matter curves,  their defining equations, the homology classes and the multiplicities
		in terms of the flux restrictions.	}
	\end{center}
	\label{SU4table}
\end{table}

 Having defined the basic ingredients of the effective model and before we proceed with the implications,
  a few comments are necessary.  We first  recall that all the matter fields are effectively representations
   of the $SU(5)\times U(1)_Z$  gauge symmetry, with $ U(1)_Z$ being the linear combination~(\ref{Zydef}) of the two $U(1)$'s
   embedded  in ${\cal E}_8$. All matter and Higgs fields in the present case  transform under the fundamental 
   representation of $SU(3)_{\perp}$. 
   In the spectral cover the latter is replaced by the Cartan subalgebra characterised by the weights $t_i$
   and hence all GUT fields will be distinguished by an index $t_i$.\footnote{This is in contrast to 
   	 $SU(5)\times SU(5)_{\perp}$ where, while the $SU(5)$ GUT 10-plets transform under $5\in SU(5)_{\perp}$
   	and carry indices $10_{t_i}$, the GUT 5-plets transform as $10\in SU(5)_{\perp}$ 
   		and therefore are denoted as $\bar 5_{t_i+t_j}$.} 
   
   In the following we will discriminate matter curves with respect to $SU(3)_{\perp}$
using the subscripts $t_i$ and, to avoid clutter in the formalism, we omit the hypercharge subscript.

The ${\cal E}_6$ Yukawa coupling ${\bf 27}^3$ implies the following generic form of $SO(10)$ 
invariants
\ba
{\cal W}_{SO(10)}&=& {\bf 16}_{t_i}{\bf 16}_{t_j}{\bf 10}_{t_k}+{\bf 10}_{t_i}{\bf 10}_{t_j}{\bf 1}_{t_k}\,,
\label{WSO10}
\ea
with $i, j, k$ such  that $t_i+t_j+t_k=0$. Under further breaking of the  symmetry down to $SU(5)\times U(1)$, 
 the couplings as well as the associated mass matrices are as follows. 
The first coupling in~(\ref{WSO10})  implies~\footnote{Bold face is 
 used for the $SO(10)$ representations, to avoid confusion with those of $SU(5)$.}
\ba
{\bf 16}_{t_i}{\bf 16}_{t_j}{\bf 10}_{t_k}&=&
\left\{\begin{array}{ll}
                        10_{t_i}^f10_{t_j}^f5_{t_k}^{h_d}&\ra m_{down}\\
                        10_{t_i}^f\bar 5_{t_j}^{h_u}\bar 5_{t_k}^f&\ra m_{up},m_{\nu}
                           \end{array}\right.\,,
                           \label{16Dec}
         \ea
while the second coupling in (\ref{WSO10}) gives
\ba
{\bf 10}_{t_i}{\bf 10}_{t_j}{\bf 1}_{t_k}&=& { 5}^{h_d}_{t_i}\,{\bar 5}^f_{t_j}\,{\bf 1}_{t_k}+
{ 5}^{h_d}_{t_j}\,{\bar 5}^f_{t_i}\,{\bf 1}_{t_k}\ra m_{\ell}\,.
\ea
 As is expected in flipped models, we observe that the up-quark masses emerge from the $SU(5)$ coupling $10\cdot\bar 5\cdot\bar 5$
	and this implies that $\lambda_t=\lambda_{\nu}$ at the GUT scale.
	The bottom quark masses are obtained from $10\cdot 10\cdot 5$, while the charged lepton masses arise from $\bar 5\cdot 5\cdot 1$.
	As a consequence, the well known relation $\lambda_{\tau}=\lambda_{b}$ at $M_{GUT}$ of standard $SU(5)$  is no longer applicable. 

The above $SO(10)$ couplings in general involve potentially dangerous terms leading to 
baryon and lepton number violation at unacceptable rates. An additional effect of the 
$U(1)$ fluxes introduced to break the symmetry however, is to `eliminate' various 
components of the decomposed representations and generate chiralities.  Keeping this 
in mind as well as the required massless spectrum of the effective model, we find that the
 massless spectrum should be arranged as follows.
 All families should be accommodated in the $27_{t_{1}}$ curve,
\ba
{\bf 27}_{t_{1}}&\ra& {\bf 16}+{\bf 10}+{\bf 1}\;\to \;
       \{10^f+\phi\}_{t_{1}}+\{\bar 5^f\}_{t_{1}}+{\bf 1}_{t_{1}}\,.
\ea
In the last step  we assume that the flux induces chirality only for the $SU(5)$ components $10^f$ 
and $\phi$ of the ${\bf 16}$-plet and only for $\bar 5^f$ of the ${\bf 10}$-plet. On the contrary,
the $\bar 5^{h_u}\in {\bf 16}$ and $5^{h_d}\in {\bf 10}$ of $SO(10)$ are not included, meaning that 
in the effective theory either they are eliminated or they appear in vector-like pairs with their
corresponding (conjugate) representations from the other matter curve. 

In a similar way for the Higgses we require:
\ba
{\bf 27}_{t_{3}}&\ra& {\bf 16}_{t_3}+{\bf 10}_{t_3}+{\bf 1}_{t_3}\;
        \ra\; \{\bar 5^{h_u}+\phi\}_{t_{3}}+\{ 5^{h_d}\}_{t_{3}}+{\bf 1}^f_{t_{3}}\,.
\ea
In the second  step we assume that along the $t_3$ curve, the flux eliminates $10_{t_3}\in {\bf 16}$ and
$\bar 5^f_{t_3}\in {\bf 10}$.
With this assignment, we readily obtain the following  $SU(5)$ invariant couplings~\footnote{In terms of
	the $U(1)_{X,X'}$ initial charges these couplings are written as follows:\\ $$10^f_{(-1,1)} 10^f_{(-1,1)} 5^{h_d}_{(2,-2)}, 
 \, 10^f_{(-1,1)} \bar 5^f_{(-2,-2)} 5^{h_u}_{(3,1)},\,10^f_{(-2,-2)} \bar 5^{h_d}_{(2,-2)} 1^{f}_{(0,4)}$$. }
\ba
{\cal W}&=& 10^f_{t_1} 10^f_{t_2} 5^{h_d}_{t_3}
+ 10^f_{t_1}\bar 5^{f}_{t_2}\bar 5^{h_u}_{t_3}
+\bar 5^{f}_{t_1}1^f_{t_2} 5^{h_d}_{t_3}\,,
\ea
which are just the mass terms for bottom, top, and Dirac neutrino and charged leptons respectively.
To obtain a light Majorana sector we may appeal to the well-known see-saw mechanism which,
in F-theory constructions  can give an ${\cal O}(10^{-1}{\rm eV})$ scale  from integrating out 
Kaluza-Klein modes~\cite{Bouchard:2009bu}.  

  We emphasize that in our framework the chiral states emerge only from the $\Sigma_{27}$ 
 matter curves.  As we have seen   in the previous section the  chirality of the 
 spectrum is ensured  by suitable restrictions of $U(1)$  fluxes  inside $SU(3)_{\perp}$. 
 We  assume that states coming from the 78 always appear  in vector-like pairs. In  section 4
 we will see how this can be realised, despite the appearance of non-trivial fluxes which will be turned on 
 along $U(1)$ factors inside $E_6$.

A possible way to eliminate the unwanted representations can be illustrated in the following simple example.
Suppose that on a suitable line bundle, choosing the topological properties as already explained above,
 we can  obtain three chiral  ${\cal E}_6$ fundamental representations,  $n_{27}-n_{\overline{27}}=3$.
Turning on an abelian flux  along $U(1)_{X'}$ we obtain the following decompositions 
\ba
{\bf 27}_{t_{1}}
=
\left\{\begin{array}{ll}{\rm Rep.}&$\;\;\;\#$\\
                        {\bf 16}_{t_1}&:\;x_1\\
                        {\bf 10}_{t_1}&:\;x_1-n_1\\
                        {\bf 1}_{t_1}(e^c)&:\;x_1+n_1
                           \end{array}\right.
                           &&
{\bf 27}_{t_{3}}
=
\left\{\begin{array}{ll}{\rm Rep.}&$\;\;\;\#$\\
                       {\bf 16}_{t_3}&:\;x_3\\
                        {\bf 10}_{t_3}&:\;x_3-n_3\\
                        {\bf 1}_{t_3}(\bar e^c)&:\;x_3+n_3
                           \end{array}\right.\,.
                           \label{10dec12}
\ea
The integers $x_{1,3}$ represent the number of ${\bf 27}_{t_1},{\bf 27}_{t_3}$ representations
and  $n_{1,3}$ are the integers associated with the flux restrictions on the corresponding matter curves.
We wish to accommodate the fermion generations along $t_1$ and for the simplest choice $x_1=3, x_3=0$ 
for example, we  guarantee the existence of  three chiral $16_{t_1}$ representations which 
contain an equal number of $10\in SU(5)$.   Furthermore, for the 
	flux we impose the restriction  $\sum_in_i=n_3+n_1=0$.   If  we  set  $0\le n_1\le x_1$, 
 we get $3-n_1\ge 0$ representations of
 ${\bf 10}_{t_1}\in SO(10)$ and $3+n_1$ singlets with the quantum numbers of $e^c$. 
 On the $t_3$ matter curve there are also $n_1$ singlet fields transforming in 
 the conjugate representation of $e^c$. Hence, 
 three of the  $ 1_{t_1}(e^c)$ singlets accommodate the $e^c, \mu^c, \tau^c$,
 while the remaining $n_1$ form  vector-like pairs  $(1_{t_1},1_{-t_3})\to ( E^c,\bar E^c)$. 
 On the contrary, for $n_1=-1,-2,-3$ there are only three singlets  $ 1_{t_i}(e^c)$ available
 which exactly match the SM states $e^c, \mu^c, \tau^c$. However, this is no longer true for
 the $10$-plets, and, as can be  readily observed  there now exist additional vector-like pairs.

We should point out that if the geometric singularity associated with the GUT symmetry 
is $SO(10)$ or higher~\cite{Beasley:2008kw}, there is no way to eliminate all the exotic states from the
spectrum of the effective model. Therefore, in this ${\cal E}_6$ construction  additional
states always  appear in the massless spectrum. Their implications for gauge coupling unification
as well as in the effective field theory models emerging from ${\cal E}_6$ group
 have been studied in~\cite{Callaghan:2013kaa}.
The rather interesting fact is that they appear in vector-like pairs so that they can 
form massive states provided there exist appropriate couplings to singlet fields acquiring vevs.

 The number of SU(5) irreps in ${\bf 16}$'s and  ${\bf 10}$'s of $SO(10)$ 
 can be similarly defined by the flux parameters along the $U(1)_X $:
\ba
{\bf 16}_{t_{i}}
=
\left\{\begin{array}{ll} 
                        10^f_{t_i}&:\;x_i\\
                        \bar 5^{h_u}_{t_1}&:\;x_i+m_i\\
                        1_{t_i}(\phi)&:\;x_i-m_i
                           \end{array}\right.\;,
                           &&
   {\bf 10}_{t_{i}}=
                           \left\{\begin{array}{ll} 
                        5^{h_d}_{t_i}&:\;x_i-n_i\\
                       \bar 5^f_{t_i}& :\;x_i-n_i+p_i\\
                           \end{array}\right.\,,
                           \label{10dec12}
         \ea
 for $i=1,3$ and the total $U(1)_X$- flux  is taken to be zero.  The flux vanishes automatically  on the   
 $16_{t_i}$ components, ($\sum_im_i+(- m_i)=0$),  which implies $p_1+p_3=0$.
 
   The  values already chosen for $x_{1,3}$ ensure that we have $3\times (10_{t_1}+e^c_{t_1})$
 multiplets. However, to obtain three complete families we need $3\times \bar 5^f_{t_i}$ 
 multiplets. In the simplest scenario we can obtain them from the $t_1$ matter curve, and so we can simply require  
 $x_1-n_1+p_1=3$ which is  satisfied for $p_1=n_1$ and $p_3=n_3=-n_1$.
 
\begin{table}
	\begin{center}
		\begin{tabular}{l|ll|lc}
			$SO(10)$& $SU(5)$& $\# $ &$SU(5)$ &$\# $\\
			\hline
			${\bf 16}$&$10_{t_1}^f$&$3$&$10_{t_3}^f$&$0$
			\\
			${\bf 10}$&$\bar 5_{t_1}^f$&$3$&$\bar 5_{t_3}^f$&$0$
			\\
			${\bf 1}$&$1_{t_1}^f(e^c)$&$3+n_1$&$1_{t_3}^f(\bar e^c)$&$-n_1$
			\\
			${\bf 10}$&$5_{t_1}^{h_d}$&$3-n_1$&$5_{t_3}^{h_d}$&$n_1$
			\\
			${\bf 16}$&$\bar 5_{t_1}^{h_u}$&$3+m_1$&$\bar 5_{t_3}^{h_u}$&$-m_1$
			\\
			${\bf 16}$&$1_{t_1}$&$3-m_1$&$1_{t_3}$&$m_1$
			\\
			${\bf 1}$&$1_{t_1-t_3}$&$n$&$1_{t_3-t_1}$&$n$\\
		\end{tabular}
		\caption{$SO(10)$ origin and multiplicities of the $SU(5)$ content. If for some choice
			of the integers $m_1, n_1$  the multiplicity	turns negative, then one obtains the conjugate representation.
			In the last row we include the ${\cal E}_6$  singlets $\theta_{13}$ and $\theta_{31}$ with (unspecified) multiplicity $n$.}
	\end{center}
	\label{B1}
\end{table}

To allow for  tree-level invariant Yukawa couplings the Higgs fields are accommodated in $5^{h_{d}}\in {\bf 10}_{t_3}$ and  
$\bar 5^{h_{u}}\in {\bf 16}_{t_3}$. In general, depending on the specific choice of the 
parameters $n_i, m_i$, there can be additional vector-like pairs of $ 5^{h_{d}}_{t_1}, 5^{h_{d}}_{t_3}$ 
and $\bar 5^{h_{u}}_{t_1},  \bar 5^{h_{u}}_{t_3}$ 5-lets but several of them
can receive masses from terms of the form $5_{-t_i}\cdot \bar 5_{t_j}\cdot 1_{t_i-t_j}$ or other available couplings.

Next we focus on the $SU(5)$ breaking. 
In the case of flipped $SU(5)$ there exist, in principle, more than one mechanism available for the
spontaneous breaking of $SU(5)$. This can happen either with  the hypercharge flux, or with the Higgs mechanism.
The hypercharge flux  breaking  of $SU(5)$ down to SM  splits the $SU(5)$ multiplets when 
its restriction  on the corresponding matter curve is non-trivial. In general, we wish to maintain  intact the 
representations accommodating the fermions, and hence we may assume that the matter curve $t_1$ is not affected.
For the case of Higgs 5-plets, this differentiates the number of doublets and triplets residing on the
same matter curve. 
\begin{table}
	\begin{center}
		\begin{tabular}{l|llc|llc}
			$SO(10)$& $SU(5)$&$U(1)_{(X,X')}$& $\# $ &$SU(5)$ &$U(1)_{(X,X')}$&$\# $\\
			\hline
			${\bf 16}$&$10_{t_1}^f$&$(-1,1)$&$3$&$10_{t_3}^f$&$-$&$0$
			\\
			${\bf 10}$&$\bar 5_{t_1}^f$&$(-2,-2)$&$3$&$\bar 5_{t_3}^f$&$-$&$0$
			\\
			${\bf 1}$&$1_{t_1}^f(e^c)$&$(0,4)$&$6$&$1_{-t_3}^f(\bar e^c)$&$(0,-4)$&$3$
			\\
			${\bf 10}$&$5_{t_1}^{h_d}$&$-$&$0$&$5_{t_3}^{h_d}$&$(2,-2)$&$3$
			\\
			${\bf 16}$&$\bar 5_{t_1}^{h_u}$&$-$&$0$&$\bar 5_{t_3}^{h_u}$&$(3,1)$&$3$
			\\
			${\bf 16}$&$1_{t_1}$&$(-5,1)$&$6$&$1_{-t_3}$&$(5,-1)$&$3$
			\\
			${\bf 1}$&$\theta_{13}$&$(0,0)$&$n$&$\theta_{31}$&$(0,0)$&$n$
			\\
		\end{tabular}
		\caption{$SO(10)$ origin and multiplicities of the $SU(5)$ content in the first example (see text).
			The $U(1)_{X}, U(1)_{X'}$ charges of the $SU(5)$ representations are also shown. }
		\label{B2}
	\end{center}
\end{table}
 We note however,  that there are strong constraints on the spectrum due to the fact that
the hypercharge flux should be globally trivial~\cite{Marsano:2010sq,Palti:2012dd,Cvetic:2012xn}.
In the present model, however, we can evade such constraints if we appeal to the ordinary
Higgs mechanism for the $SU(5)$ breaking. Indeed,  since we deal with flipped $SU(5)$~\cite{Barr:1981qv},
we can use the  $10_H+\ov{10}_{\bar H}$ Higgs-pair to break the symmetry\cite{Antoniadis:1987dx}. 
In this case we assume that the flux also vanishes along the Higgs and matter curves.  
Such vector-like pairs are  then available, and to ensure that these are  massless, we 
require~\cite{Donagi:2008ca}  topological properties such that $h^0(\Sigma, K^{1/2}_{\Sigma})=0$, 
where $\Sigma$ is the corresponding matter curve and $K_{\Sigma}$ the canonical bundle. 

 For the implementation of this mechanism we will consider a variant of the minimal spectrum 
	presented in Table~\ref{B2B}.	Assuming the existence of one massless vector-like pair $10^H_{t_1}+\overline{10}^H_{-t_1}$
	with content denoted as $10^H=(Q_H,\nu_H^c, D^c_H)$ and $\overline{10}^H=(\bar Q_H,\bar\nu_H^c, \bar D^c_H )$,
	 the following couplings are generated
	\be 
	{\cal W}_H \supset 10^H_{t_1}10^H_{t_1}5^h_{t_3}+\overline{10}^H_{-t_1}\overline{10}^H_{-t_1}\bar{5}^h_{-t_3}
	\to \nu_H^c D^c_HD + \bar\nu_H^c \bar D^c_H \bar D \,.
	\ee 
	As the neutral singlets $\nu_H^c ,\bar\nu_H^c$ acquire vevs, the associated triplet pairs become massive and only the corresponding
	doublet fields $h_u, h_d$ survive ($Q_H, \bar Q_H$ are  the longitudinal components of gauge bosons). Clearly,
	this doublet-triplet splitting mechanism affects only one pair of Higgs 5-plets, leaving all others intact. 	
	Thus, we end up  with the MSSM spectrum and only vector-like 5-plets.

\begin{figure}[!ht]
	\centering    
	\includegraphics{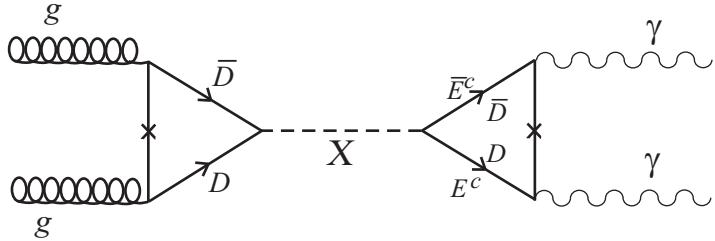}       
	\caption{Vector-like  representations $E^c+\overline{E^c}$ and $(D+\bar D)\in (5+\bar 5)$ 
		of $SU(5)$ 	contribute to the diphoton signal.	The 750 GeV resonance 
		is associated with the MSSM singlets of the model.}
	\label{ggXgg}
\end{figure}
Finally let us comment on two important issues, namely  the $\mu$-term
and gauge coupling unification. As we have seen, the up- and down-Higgs
doublets are accommodated in $\bar 5_{-2}\in {\bf 16}_{1}$ and $5_{2}\in {\bf 10}_{-2}$ 
representations of $SU(5)\times U(1)_Z\in SO(10)\times U(1)_{X'}$. As can be observed,  
the $U(1)_Z$ symmetry alone, could not solve the naturalness problem, 
since it would allow a term $\bar 5_{-2}5_{2}$ with a mass of the order of the GUT scale.  
Notice however, that both Higgs 5-plets emerge from the ${\bf 27}_{t_3}$-matter curve,
and therefore this term does not explicitly  appear in the Lagrangian because of a non-zero
$U(1)_q$ charge (equal to $2t_3$ units), which cannot be cancelled by a tree-level term
containing the available neutral singlets capable of developing vevs. Therefore,
we can safely appeal to the Giudice-Masiero~\cite{Giudice:1988yz} F-term mechanism for a
 natural solution of the $\mu$-problem.
For the same reason, it is not possible to have a direct tree-level coupling for the 
colour triplet-antitriplet pair of the Higgs 5-plets, thus avoiding  the rapid proton-decay problem.

Next, we will provide a brief account of gauge coupling unification which, 
in F-theory constructions, is a rather complicated issue. We have seen  in the present $E_6$ 
construction that the massless spectrum contains the MSSM fields, four complete $5+\bar 5$ pairs
and  $E^c+\bar E^c$ pairs with the quantum numbers of the right-handed electron. 
The $5+\bar 5$ pairs modify only the common value of the gauge coupling at $M_{GUT}$
but do not affect  gauge coupling unification. This  no longer holds because of  the 
additional $E^c+\bar E^c$  fields, but  in F-constructions this is not the whole picture.

  Indeed, we first point out that the massive Kaluza-Klein modes affect  gauge coupling unification
  since their mass scale (associated also with the right-handed neutrino mass) is comparable to the 
  GUT scale~\cite{Bouchard:2009bu}.
 Moreover,
in F-theory derived GUT models  there are many massive states charged under the Standard
 Model gauge group which  split the values of the gauge couplings at the GUT scale. 
 Hence, 
 in general we do not expect precise unification~\cite{Donagi:2008kj,Leontaris:2011tw}. Furthermore, the $E_6$  
 gauge symmetry breaking via  fluxes is another source of  gauge couplings' splitting~\cite{Blumenhagen:2008aw,Leontaris:2009wi}. 
 In certain cases, it can be shown that several combinations of additional matter can compensate 
 for this splittings leading to a consistent unification scenario.
 Finally, to keep the hypercharge interaction within the perturbative regime 
below the GUT scale,  some of the extra pairs of 5-plets and $E^c+E$ singlets  
 should receive masses well above the TeV region. This can be achieved if suitable singlet
 fields acquire apropriate non-zero vevs. A possible way to provide large  mass to these fields 
 is by a fourth order non-renormalizable (NR) term of the form 
$(\theta_{13}+\theta_{1j}\theta_{j3}) \rangle \langle1_{t_1}\rangle  5_{t_3}^{h_d}\bar 5_{t_3}^{h_u},$
 where   either the $ \theta_{13}$  singlet  or the combination $\theta_{1j}\theta_{j3}, j=4,5$ acquire non-zero vev.
A  complete account of  these issues, however, is beyond the goals of the present work.

\section{Diphoton emission}

As pointed out above, a particularly  interesting fact in these F-theory based
constructions is the presence of exotic matter in  the low energy effective theory. 
Since matter in these models has a geometric origin,  the precise massless content  depends 
on the specific choice of the compactification manifold and the fluxes are parametrised in terms 
of a few integer parameters.

 In the present example this  dependence is shown in Table~\ref{B1}. 
 Two specific choices lead to the spectra given below. Starting  with a minimal
 case, we choose $n_1=-m_1=3$ and the resulting spectrum is shown in Table~\ref{B2}.
There are three additional
 vector-like charged leptons with the quantum numbers of $E^c+\bar E^c$ which can contribute to the
 decay of the singlet scalar to two photons as shown in the right loop of figure~\ref{ggXgg}.

However,  this minimal choice of parameters does not provide  vector-like pairs of coloured particles
to mediate the $X$-production from gluon fusion, because the couplings of these states to an appropriate
singlet are suppressed. 
 This can be achieved if we instead choose  $n_1=-m_1=4$.
The resulting spectrum is shown in Table~\ref{B2B}.  There are now four vector-like pairs of $E^c+\bar E^c$
 as well as the additional pairs  
$ 5^{h_d}_{t_3}+ \bar 5^{\bar h_d}_{-t_1}$  and $\bar 5^{\bar h_u}_{-t_1}+\bar 5^{h_u}_{t_3}$.
\begin{table}
	\begin{center}
		\begin{tabular}{l|llc|llc}
			$SO(10)$& $SU(5)$&$U(1)_{(X,X')}$& $\# $ &$SU(5)$ &$U(1)_{(X,X')}$&$\# $\\
			\hline
			${\bf 16}$&$10_{t_1}^f$&$(-1,1)$&$3$&$10_{t_3}^f$&$-$&$0$
			\\
			${\bf 10}$&$\bar 5_{t_1}^f$&$(-2,-2)$&$3$&$\bar 5_{t_3}^f$&$-$&$0$
			\\
			${\bf 1}$&$1_{t_1}^f(e^c)$&$(0,4)$&$7$&$1_{-t_3}^f(\bar e^c)$&$(0,-4)$&$4$
			\\
			${\bf 10}$&$\bar 5_{-t_1}^{\bar h_d}$&$(-2,2)$&$1$&$5_{t_3}^{h_d}$&$(2,-2)$&$4$
			\\
			${\bf 16}$&$ 5_{-t_1}^{\bar h_u}$&$(-3,-1)$&$1$&$\bar 5_{t_3}^{h_u}$&$(3,1)$&$4$
			\\
			${\bf 16}$&$1_{t_1}$&$(-5,1)$&$7$&$1_{-t_3}$&$(5,-1)$&$4$
			\\
			${\bf 1}$&$\theta_{13}$&$(0,0)$&$n$&$\theta_{31}$&$(0,0)$&$n$
			\\
		\end{tabular}
		\caption{The $SU(5)$ representations, their $U(1)_{X}, U(1)_{X'}$ charges 
			and the corresponding multiplicities for the second case (see text). }
		\label{B2B}
	\end{center}
\end{table}
As a result, they  generate the superpotential terms
\be  
 \bar 5^{\bar h_d}_{-t_1}\cdot 5^{h_d}_{t_3}  \cdot \theta_{13}+ 5^{\bar h_u}_{-t_1}\cdot\bar 5^{h_u}_{t_3}  \cdot \theta_{13} + 
(\bar E^c)_{-t_3}(E^c)_{t_1}\cdot \theta_{31} \label{t13}\,.
\ee 
We identify the $750$ GeV resonance $X$ with  the singlet $\theta_{13}$  which has the appopriate couplings 
to  give rise to the diphoton diagram shown in figure~\ref{ggXgg}.   Note that a mixing term $\theta_{13}\theta_{31}$
 would allow an additional channel for the resonance to decay into diphotons through the last coupling in equation~(\ref{t13}). 
After supersymmetry breaking  in the presence of the scalar $X$, the effective Lagrangian contains the terms
\ba 
{\cal L}&=& \lambda_D\bar DDX+\lambda_E\bar E^cE^cX+\frac 12M_X^2XX+A_DX\tilde D^*\tilde D+
A_EX\tilde E^{c*}\tilde E^c+\cdots\,.
\ea
We assume here that the   singlet field receives a soft mass $M_X$ of the
order  of the  SUSY breaking  scale, $\lambda_{f}, f=D,E$ are  Yukawa couplings of order one, and $A_{D,E}$
are the trilinear scalar parameters.  For a pseudoscalar interaction  we should replace the Yukawa  
coupling according to $\lambda_f\to i\gamma_5 \lambda_{5f}$.

Next we provide an estimate of the contributions of the above exotics to the diphoton excess.  
We assume that the production  mechanism of the scalar resonance is mainly from  gluon fusion,
 mediated by loops of the colour triplets, while its decay  is mediated by triplets and ($E^c, \bar E^c$)-pairs 
 as shown in the figure. The cross section for the scalar mediated process is 
 \[ \sigma(gg\to X\to\gamma\gamma)
  =\frac{1}{M_X\cdot\Gamma\cdot s}C_{gg}\Gamma(X\to gg)\Gamma(X\to\gamma\gamma)\,,\]
where $\Gamma, \sqrt s$ are the total width and the center of mass energy ($\sqrt s=13$TeV) 
respectively, and $C_{gg}$ is 
 the parton integral~\cite{Martin:2009iq}
\[  C_{gg}=\int_{M_X/s}^1 f_g(x)f_g( \frac{{\tiny M_X}}{sx})\frac{dx}{x}\,,   \]
where $f_g(x)$ is the function representing the gluon distribution  inside the proton. 
The integral is computed using MSTW2008NNLO~\cite{Martin:2009iq}
and its numerical value at $13$ TeV is estimated~\cite{Franceschini:2015kwy} to be $C_{gg}=2137$.
The partial widths $ \Gamma(X\to gg),\Gamma(X\to\gamma\gamma)$ 
from loops involving fermions and scalars  are given by~\cite{Djouadi:2005gi}(see also\cite{Franceschini:2015kwy})
\ba 
\frac{\Gamma(X\to gg)}{M_X}&=&\frac{\alpha_3^2}{2\pi^3} \left|\sum_fC_{r_f}\sqrt{\tau_f}\lambda_fS(\tau_f)
           +\sum_s C_{r_s}\frac{A_s}{2M_X} P(\tau_s)\right|^2\,,\\
 \frac{\Gamma(X\to \gamma\gamma)}{M_X}&=&\frac{\alpha^2}{16\pi^3} \left|\sum_fd_{r_f}q_f^2\sqrt{\tau_f}\lambda_fS(\tau_f)
            +\sum_s d_{r_s}q_s^2\frac{A_s}{2M_X} P(\tau_s)\right|^2\,,          
\ea 
where $C_r$ is the Dynkin index of the colour representation $(C_r=3$ for the triplet), $d_r$ is its dimension, 
$q_s$ the charge and $\sqrt{\tau_{a}}=\frac{2 m_{a}}{M_X}$, with $a=f,s$ for the fermion and scalar  masses respectively.
The functions $ S(\tau), P(\tau)$ are 
\[ S(\tau) = 1+(1-\tau) f(\tau),\; P(\tau)=\tau f(\tau)-1\,,\]
where~\cite{Shifman:1979eb}
\ba 
f(\tau)&=&\left\{\begin{array}{ll}
                  \arctan^2\frac{1}{\sqrt{\tau-1}}&\tau>1   \\
                  -\frac 14\left(\log\frac{1+\sqrt{1-\tau}}{1-\sqrt{1-\tau}}-i\pi\right)^2;&\tau\le 1
                  \end{array}
\right.\,.
\ea 
For the pseudoscalar contribution, in the above formulae we make the replacements
 $S(\tau_f)\to f(\tau_f), P(\tau_s)\to 0$ and $\lambda_f \to \lambda_{5f}$~\cite{Djouadi:2005gi,Franceschini:2015kwy}. 

\begin{figure}[!ht]
	\centering       
		\includegraphics[scale=.53,angle=0]{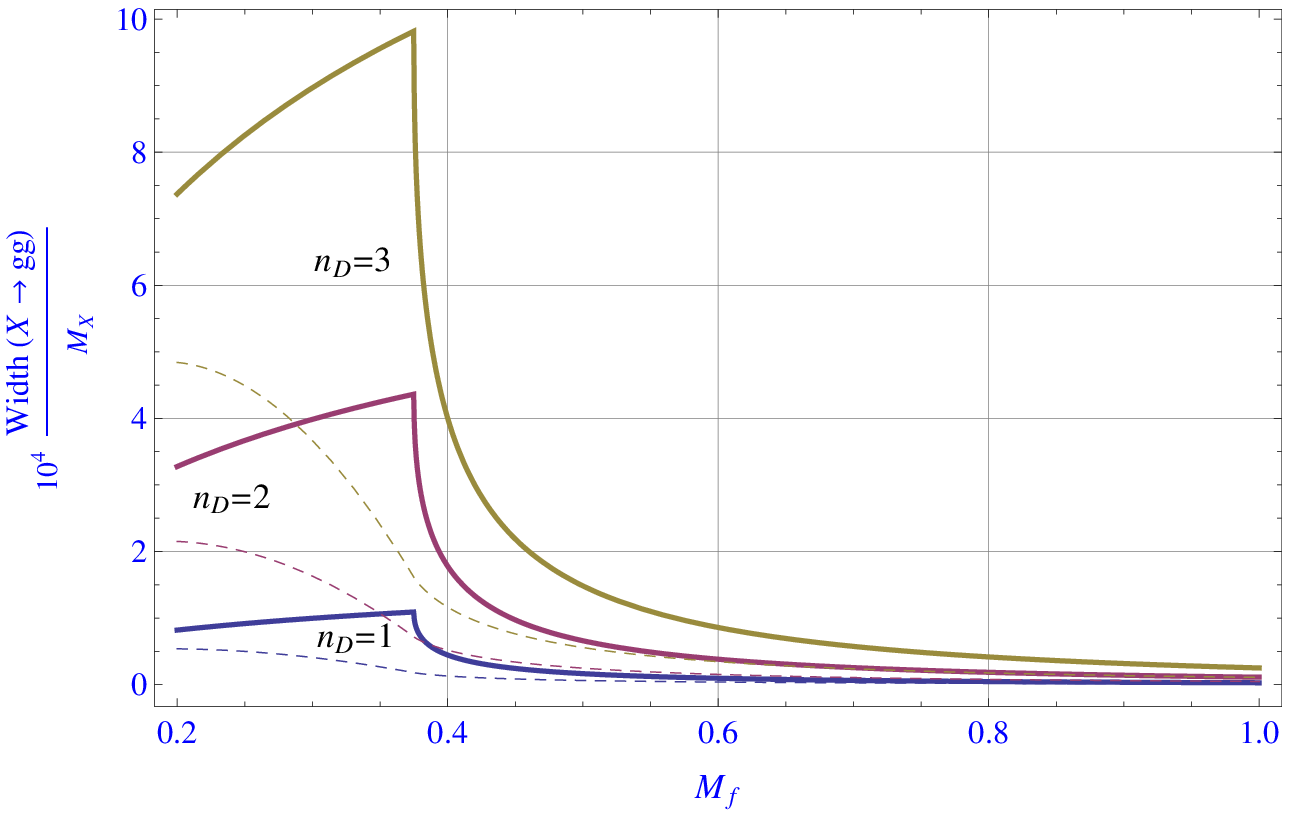}\; 
		\includegraphics[scale=.53,angle=0]{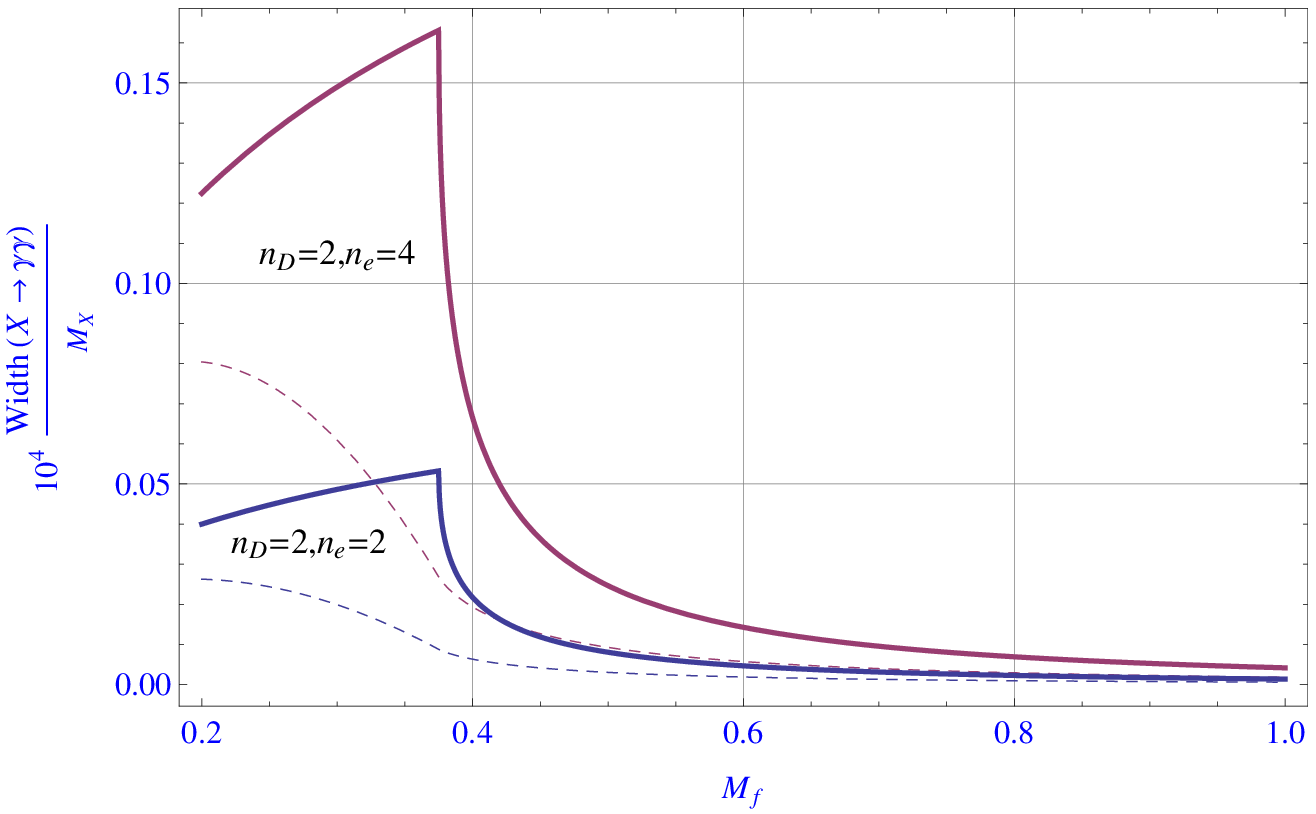} 
	\caption{Plots of the ratios $\frac{\Gamma_{gg}}{M_X}$ and $\frac{\Gamma_{\gamma\gamma}}{M_X}$ 
	as a function of the masses of the fermion-pairs circulating in the loops of figure~\ref{ggXgg}.}
	\label{XggXgamma}
\end{figure}

For a numerical application,  we first consider the existence of only one singlet field $X$ with mass $M_X=750$ GeV
and, for the sake of simplicity, we take a common mass for the various fermion-pairs contributing in the loops.
 Since the scalar components
are expected to be much heavier than the fermions, at this level of approximation their contributions are ignored. 
In figure~\ref{XggXgamma} we plot the widths $\Gamma_{gg}$ and $\Gamma_{\gamma\gamma}$ as a function of 
the mass of the fermion-pairs for two sets of fermion multiplicities for the scalar as well as the pseudoscalar case.
If we ignore the large width suggested by the ATLAS data, we observe that there are regions of the fermion mass range 
where $\Gamma_{gg}\sim \,{\rm few}\; 10^{-4}M_X$ and  $\Gamma_{\gamma\gamma}\sim \, {\rm few}\;  10^{-6}M_X$, which are 
sufficient to interpret the data. 
We note in passing that a large decay width allows the exciting possibility of other decay channels including dark matter.
 In general, however, we expect more than one singlet field with approximately degenerate
masses, so that the ATLAS large width could be explained as an unresolved resonance. Another possibility is to invoke additional
couplings in the superpotential such as $XH_uH_d$ which permit  the resonance to decay into Higgsinos, if kinematically possible,
or SM Higgs via the soft trilinear terms. 

\subsection{Bulk matter}
 
	Before closing, we would like to make a final comment on the possible existence of   additional `exotic' matter interactions.
	As we have pointed out, exotic matter arises from the decomposition ${\bf 78}\ra {\bf  45}_0+{\bf 16}_{-3}+{\bf \ov{16}}_3+{\bf 1}_0$
	with respect to $SO(10)\times U(1)_{X'}$  (the indices now  refer to $U(1)_{X'}$).
	We recall that in the twisted model the SM states are in ${\bf 16}_{-1}$ while  bulk  
	states are the ${\bf 16}_{-3}$, and as such they have exotic charges.  
	Such states could pick up masses at a high scale. in case some of them 
	remain light. 
	 Due to their large  $Y$-hypercharge,  they  can in principle make  a significant 
	contribution to the production and decay of the resonance.
	As can be observed, all these states come in vector-like pairs,	
	and therefore a  possible coupling that could  make them massive is 	
	\be 
	{\bf 78}\cdot {\bf 78}\cdot {\bf 1} \to M_{Q'} Q'\bar Q' +\cdots  \,.  \label{M78}
	\ee 
	Since these states carry non-zero `charges'  under the three $U(1)$'s, in principle, non-trivial
	fluxes might lead to additional chiral states.  Nevertheless, a solution to this problem is feasible
	if certain topological properties are assumed. 

 Indeed, we first recall that the number of states is given by  the Euler character $\chi$.
	If  $\tau^*$ is the dual representation of $\tau$, ${\cal T}$ is the bundle transforming in the
	representation $T$,  the net number of chiral  minus anti-chiral
	states is given in terms of the formula~\cite{Beasley:2008dc},
	$$ n_{\tau}-n_{\tau^*}=\chi(S,{\cal T}_j^*)-\chi(S,{\cal T}_j)$$
	where we assume $S$  to be a  del Pezzo surface associated with the gauge group  $G_S$.
	If we designate with ${\cal L}_j$ a line bundle over $S$, the Euler character is 
	\ba
	\chi(S,{\cal L}_j)&=&1+\frac 12 c_1({\cal L}_j)\cdot  c_1({\cal L}_j)+\frac 12 c_1({\cal L}_j)\cdot c_1(S)\nn\\
	\chi(S,{\cal L}_j^*)&=&1+\frac 12 c_1({\cal L}_j)\cdot  c_1({\cal L}_j)-\frac 12 c_1({\cal L}_j)\cdot c_1(S),
	\ea
	so that the difference counting the number of chiral states is 
	\be
	\chi(S,{\cal L}_j^*)-\chi(S,{\cal L}_j)=- c_1({\cal L}_j)\cdot c_1(S)\,\cdot\label{nochiral}
	\ee 
	We can ensure the vector-like nature of the corresponding states by simply demanding
	\be 
	c_1({\cal L}_j)\cdot c_1(S)=0\label{cond4vpairs}
	\ee 
	for the particular line bundle. 
	
	Focusing   now on the  $E_6$ case, recall that under the successive breaking we have considered
	\[  E_6\supset SO(10)\times U(1)_X\supset SU(5)\times U(1)_{X}\times U(1)_{X'} \supset SU(3)\times SU(2)\times U(1)_Y\times U(1)_{X}\times U(1)_{X'} \] 
	while the quantum numbers of the bulk states  are
	 \ba 
	 78&\to& 
	 (1,1)_{(0,0,0)}+\left\{(1,1)_{(0,0,0)}+(1,1)_{(0,0,0)}+(8,1)_{(0,0,0)}+(1,3)_{(0,0,0)}+(3,2)_{(-5,0,0)}+(\bar 3,2)_{(5,0,0)}\right.\nonumber\\
	 &+&\left.(3,2)_{(1,4,0)}+(\bar 3,2)_{(-1,-4,0)}+(\bar 3,1)_{(-4,4,0)}+(3,1)_{(4,-4,0)}+(1,1)_{(6,4,0)}+(1,1)_{(-6,-4,0)}\right\}\nn\\
	 &+&\left\{ (1,1)_{(0,-5,-3)}+(\bar 3,1)_{(2,3,-3)}+(1,2)_{(-3,3,-3)}+(1,1)_{(6,-1,-3)}+(3,2)_{(1,-1,-3)}+(\bar 3,1)_{(-4,-1,-3)}\right\}\nn\\
	 &+&\left\{ (1,1)_{(0,5,3)}+(3,1)_{(-2,-3,3)}+(1,2)_{(3,-3,3)}+(1,1)_{(-6,1,3)}+(\bar 3,2)_{(-1,1,3)}+(3,1)_{(4,1,3)}\right\}\nn
	 \ea 	 
	We can express all the exotics  obtained from the decomposition of the $E_6$-adjoint ${\bf 78}$
	in terms of the following three line bundles:
	\begin{equation}
	{\cal{L}}_1 = (5,0,0), \, \, \, {\cal{L}}_2 = (1,4,0), \, \, \,{ \cal{L}}_3 = (1,-1,-3) \label{Bundles1}
	\end{equation}
	It can be shown that by imposing relations analogous to (\ref{cond4vpairs}) for the three line bundles, all exotic states appear in vector-like pairs and
	hence, no chiral matter arises from the bulk modes. Moreover, in the minimal case the extra states emerging from~${\bf 78}$
	can assemble in a $5_{ex}+\bar 5_{ex}$ pair 
	\[\bar  5_{ex}=(\bar 3,1)_{(2,3,-3)}+(1,2)_{(-3,3,-3)},\;\; 5_{ex} =(3,1)_{(-2,-3,3)}+(1,2)_{(3,-3,3)}\,.\]
As already stated, these  can receive a large mass from terms such as~(\ref{M78}),	so that gauge coupling unification
	is not affected.

{\bf Note Added}:

{ After this paper was submitted for publication the ATLAS and CMS experiments have reported results based on updated analysis including data collected during 2016. This data does not support the presence of the 750 GeV resonance previously reported in 2015 and in Moriond 2016.
We should emphasize that our string inspired $E_6$ model predicts the existence of the diphoton resonance
as well as vectorlike fields. Hopefully, some of these states can be discovered at the LHC and future colliders.
}

\newpage

\section{Conclusions}

In this work we have constructed a flipped $SO(10)\times U(1)$ model fully 
embedded in an $E_6$ GUT symmetry within an F-theory context.  We introduced abelian
fluxes along $U(1)$'s inside ${\cal E}_6$  to realise the symmetry breaking 
and  generate the chiral families in the  low energy spectrum of the effective theory.
We have presented simple cases that contain the three chiral families of quarks and leptons.
Furthermore, motivated by the 750 GeV diphoton resonance reported by the ATLAS and CMS
experiments, we  have given examples where 	the low energy spectrum consists of vector-like fields 
with a variety of MSSM quantum  numbers containing both coloured and leptonic states, as well as gauge singlets.
The flipped SO(10) model yields several vector-like $(E^c, \bar E^c)$-pairs whose presence could
 enhance the diphoton decay mode of the scalar resonance.

\vspace{1cm} 
{\bf Acknowledgements}. {\it G.K.L. would like to thank the Physics and Astronomy Department and Bartol Research
Institute of the University of Delaware for kind hospitality. Q.S. is supported in part by the DOE grant ``DOE-SC-0013880''.}

\end{document}